\documentclass[fleqn,usenatbib]{mnras}

\usepackage{newtxtext,newtxmath}

\usepackage[T1]{fontenc}

\DeclareRobustCommand{\VAN}[3]{#2}
\let\VANthebibliography\thebibliography
\def\thebibliography{\DeclareRobustCommand{\VAN}[3]{##3}\VANthebibliography}

\usepackage{graphicx}	
\usepackage{amsmath}	
\usepackage{multirow}
\usepackage{stfloats}
\usepackage{float}

\title[Environmental effects on satellite galaxies]{Environmental effects on satellite galaxies from the perspective of cold gas}

\author[ Hongxing Chen et al.]{
Hongxing Chen$^{1,2}$\thanks{E-mail: hchen792@wisc.edu},
Lizhi Xie $^{3,4}$\thanks{E-mail: xielizhi.1988@gmail.com},
Jie Wang$^{1}$,
Wenkai Hu$^{5}$, 
Gabriella De Lucia$^{4,6}$,
Fabio Fontanot$^{4,6}$, 
\newauthor
Michaela Hirschamnn$^{4,7}$
\\
$^{1}$National Astronomical Observatories, Chinese Academy of Sciences, Beijing 100101, China \\
$^{2}$Department of Astronomy, University of Wisconsin–Madison, Madison, WI 53706, USA \\
$^{3}$Tianjin Astrophysics Center, Tianjin Normal University, Binshuixidao 393, 300384 Tianjin, China \\
$^{4}$INAF – Astronomical Observatory of Trieste, via G.B. Tiepolo 11, I-34143 Trieste, Italy \\
$^{5}$Aix Marseille Université, CNRS, LAM (Laboratoire d'Astrophysique de Marseille), F-13388 Marseille, France \\
$^{6}$IFPU – Institute for Fundamental Physics of the Universe, via Beirut 2, I-34151 Trieste, Italy \\
$^{7}$ Institute for Physics, Laboratory for Galaxy Evolution and Spectral modelling, Ecole Polytechnique Federale de Lausanne,\\
Observatoire de Sauverny, Chemin Pegasi 51, 1290 Versoix, Switzerland
}

\date{Accepted XXX. Received YYY; in original form ZZZ}

\pubyear{2023}

\begin{document}
\label{firstpage}
\pagerange{\pageref{firstpage}--\pageref{lastpage}}

\maketitle

\begin{abstract}

Environment plays a pivotal role in shaping the evolution of satellite galaxies. Analyzing the properties related to the cold gas phase of satellites provides insights into unravelling the complexity of environmental effects. We use the hydro-dynamical simulations Illustris TNG and Eagle, and the semi-analytic models (SAMs) GAEA and L-Galaxies, in comparison with recent observations from the Westerbork Synthesis Radio Telescope (WSRT), to investigate the properties of satellite galaxies hosted by halos with mass $M_{200}>10^{12.8}M_\odot$, and within projected regions $\le 1.1$ virial radius $R_{200}$. Generally, satellite galaxies selected from semi-analytic models have more HI than those selected from hydro-dynamical simulations across all projected radii, e.g. more than 30\% of satellites in the two hydro-simulations are HI depleted, while this fraction is almost zero in SAMs. Furthermore, both hydro-dynamical simulations and SAMs reproduce the observed decrease of HI content and specific star-formation rate (sSFR) towards the halo centre. However, the trend is steeper in two hydro-dynamical simulations TNG and EAGLE, resulting in a better agreement with the observational data, especially in more massive halos. By comparing the two version of GAEA, we find that the inclusion of ram-pressure stripping of cold gas significantly improves the predictions on HI fractions. The refined hot gas stripping method employed in one of the two L-Galaxies models also yields improved results.

\end{abstract}

\begin{keywords}
galaxies: evolution -- methods: numerical
\end{keywords}



\section{Introduction}
 
Interactions with environment influence the evolution of a galaxy. In dense environments, galaxies transform from star-forming to passive more efficiently, causing them to be redder than their counterparts in low-density regions \citep{dressler1984,peng2010}. 
Environmental effects also play a significant role in the evolution of satellite galaxies\citep{vandenBosch2008importance,  Knobel2013color}, that are hosted by halos more massive than field central galaxies of similar stellar mass. 
Notably, the quenched fraction of satellite galaxies is significantly higher than that of central galaxies at fixed stellar mass\citep{Wetzel2012evolution}. Furthermore, as these satellite galaxies approach the centre of their host halos, their quenched fraction increases \citep{Bluck2016galactic, Wang2018dearth, hu2021atomic}.

Other properties are also found to depend on the environments. 
Observational results show that galaxies residing in the denser regions, more massive halos, or in the inner regions of halos, tend to have less HI gas \citep{Gavazzi2006observations, Cortese2011effect, Fabello2012alfalfa, Brown2017stripping}. 
Additionally, the scaling relation between HI and stellar mass is found to be correlated with the colours (g-r) or (NUV-r) \citep{Cortese2011effect, Zu2020gascontent}, and reveals that red galaxies typically exhibit a lower HI fraction compared to their blue counterparts.  
Moreover, among star-forming galaxies, an anti-correlation has been found between HI mass and gas-phase metallicity at fixed stellar mass \citep{Bothwell2013fundamental, Hughes2013coldgas, Brown2018atomic}. 
The evolution of physical properties of galaxies can be used as diagnostics of quenching mechanisms. For instance, \citet{peng2015} suggested that strangulation would cause an increase in stellar metallicity for satellite galaxies. The rapid gas removal can be ascribed to ram-pressure stripping (RPS) of cold gas \citep{Fabello2012alfalfa, Vollmer2012Virgo, Odekon2016ALFALFA}. 
Observed phenomena including elongated multi-phase gas tails and perturbed morphology of gas disks are inevitably attributed to the effect of ram-pressure stripping\citep[][and references therein]{boselli2022}.

Generally, environmental effects include: a) starvation and strangulation, that suppresses the replenishment of  the interstellar medium (ISM) by halting accretion of primordial gas and stripping off the warm/hot circumgalactic medium (CGM), \citep{Larson1980evolution, Balogh2000photometry}; b) galaxy mergers, that can  ignite star-bursts and consume most of the ISM\citep{Cox2006feedback, Hopkins2013starformation}; c) stripping of ISM, i.e. direct removal of the ISM by RPS \citep{Gunn1972infall} and tidal stripping \citep{Dressler1983spectroscopy}. These processes are stronger in more massive halos\citep{xie2020influence}. In addition, internal mechanisms, like stellar winds and supernova explosions or active galactic nuclei (AGN) feedback \citep{bahe2015, zinger2020}, can heat or eject gas and make it more likely to be stripped off. Therefore, these feedback processes would be more effective in dense environments. 

Quantifying the relative contributions of different processes is challenging. All the above mechanisms are expected to be involved in the quenching process. A satellite galaxy could experience a complex environmental history, with different physical mechanisms playing the dominant role at different stages of its lifetime \citep{delucia2012, marasco2016environmental}. The impact of these mechanisms should be considered throughout the entire evolution history of galaxies. 
From the observational point of view, it is impractical to trace a galaxy's evolution and pinpoint the time intervals when different environmental processes are at play. Yet, the radial distribution of galaxies offers a glimpse of the evolution of satellite galaxies falling into massive halos\citep{woo2013}. 
Interpreting such trends is intricate, as the dependence on radial distance is a combined effect of environment density and accretion time \citep{gao2004, delucia2012, vandenBosch2016}.
Theoretical models of galaxy formation, fully embedded in a cosmological context, are necessary to disentangle these two contributions, as well as to quantify the impact of projection effects. 

Numerical tools have been extensively used and developed to improve the modelling of satellite galaxies. A consensus has emerged regarding the dominant quenching mechanisms: the feedback from the central massive black hole to quench massive galaxies \citep{Croton2006agn, somerville2015, bechmann2017, Donnari2021quenched1},  environmental effects to quench lower-mass galaxies \citep{henriques2015galaxy,stevens2017,hough2023}, and RPS to remove cold gas\citep{Bruggen2008Ram, Stevens2019}. 
Many works have been dedicated to comparing different hydro-dynamical simulations and SAMs \citep{Brown2017stripping, Shi2022, kukstas2023}. 
The modelling adopted, however, can vary significantly.  
Semi-analytic models rely on analytic equations to deal with environmental effects, while these are automatically included in hydro-dynamical simulations \citep{steinhauser2016, kulier2023}. 
While most SAMs only consider environmental effects on satellite galaxies, they overlook the impacts on central galaxies, which can also undergo significant gas stripping, as modelled by \cite{ayromlou2019new}. \citet{Wang2018dearth} argued that this assumption leads to an overestimation of the difference between central and satellite galaxies. 
SAMs that have implemented both environmental effects on hot and cold gas find good consistency with observed quenched fractions \citep{Cora2018SAM, xie2020influence}. In contrast, hydro-dynamical simulations face challenges in reproducing quenched fractions of low-mass galaxies within cluster halos \citep{bahe2017origin, Donnari2019}.
Hydro-dynamical simulations offer more natural approaches to treat environmental effects, though medium/large cosmological simulations lack the accuracy to account for the details\citep{tonnesen2021}. Moreover, the limited resolution could cause spurious numerical effects, i.e. artificially increase tidal effects and ram-pressure. 
In hydro-simulations where centrals and satellites are influenced by dense environments equally, the stripping of gas contents starts beyond the virial radius for galaxies falling into cluster halos \citep{Zinger2018}. The star formation rates and gas contents of satellite galaxies are suppressed at larger halo-centric distances in the framework of hydro-simulations with respect to semi-analytic models \citep{ayromlou2021comparing}.

In this work, we make use of the observational data sample from  \cite{hu2021atomic} to examine satellite galaxy properties as a function of distance to the host halo centre. We then conduct a comparative analysis involving several state-of-the-art hydro-dynamical simulations and semi-analytic models. Our primary goal is to assess the extent to which current theoretical models can accurately reproduce the observed trends. 
Additionally, we also divide our samples into two groups based on different halo masses, which enables us to explore the influence of halo mass dependence on the radial distribution of satellite galaxies. Our analysis is confined to the region within  1.1 $R_{200}$, thereby minimizing effects from nearby neighbour halos and concentrating solely on the influence of the host halo environment.

This paper is organized as follows. Section~\ref{sec:obsd} presents the observational data we use. In Section~\ref{sec:simd}, we give a detailed description of simulation data, galaxy properties needed in the following discussion, and how we select our galaxy samples in order to have a fair comparison with observation. Key results are discussed in Section~\ref{sec:re}, and Section~\ref{sec:cc} presents our discussion and conclusions.


\section{Observational data}\label{sec:obsd}

As described in \cite{Hu2019HI} and \cite{hu2021atomic}, original HI data come from the Westerbork Synthesis Radio Telescope (WSRT). In total, 351 hours of telescope time were used to integrate thirty-six individual pointings in the Sloan Digital Sky Survey (SDSS) South Galactic Cap region ($21h<$ RA $< 2h$ and $10^{\circ}<$ Dec $<16^{\circ}$). Data from one pointing were discarded due to bad quality. The half-power beam width (HPBW) of the WSRT is 35 arcmins, and the average synthesized beam size is $108''\times 22''$. The radio astronomy data reduction package {\sc miriad} \citep{Sault1995miriad} was used to reduce and calibrate the data. The reduced data cubes have a size of $601\times 601$ pixels with a pixel size of $3''\times 3''$, and a frequency consisting of $8\times 20$ MHz bands with a resolution of 0.15625 MHz and an overlap of 3 MHz for each band. The final data have an overall frequency ranging from 1.406 GHz to 1.268 GHz, which corresponds to a redshift range of $0.01<z<0.12$.  Considering stronger radio frequency interference (RFI) at high redshift, the upper limit is hereafter set to $z=0.11$.

The SDSS DR7 optical catalogue \citep{Abazajian2009} is then utilized to get accurate redshifts and spatial positions of galaxies covered by radio data for the subsequent stacking procedure. By cross-matching the radio data to the SDSS catalogue, a sample of 1895 galaxies in the redshift bin $0.01<z<0.11$ is obtained. This sample is also constrained by the SDSS magnitude limit, with r-band magnitude of galaxies $M_r<17.7$. The stacking techniques can then be applied to the radio data and a detailed description is given in \cite{Hu2019HI}. The main procedure includes extracting the radio spectra of galaxies, removing the residual continuum emission and de-redshifting the spectra, conserving HI flux density, deriving mass spectra and weights, correcting for beam confusion, and finally obtaining the weighted average of the mass spectra. The integrated HI mass of a stack is calculated by integrating the mass spectrum in the frequency range [$\nu_{0}-\Delta \nu$, $\nu_{0}+\Delta \nu$], where $\nu_{0}$ is 1420.406 MHz and $\Delta \nu = 1.5$ MHz. 

Group classification is obtained from the group catalogue of \cite{yang2012evolution} based on the SDSS DR7 main galaxy sample in the redshift range $0.01 < z < 0.2$. This catalogue uses iterative methods to identify groups: firstly, galaxies are assigned to tentative groups using the Friends of Friends \citep[FOF; ][]{Davis1985} algorithm; then, the characteristic luminosity of each tentative group is determined; next, the mass, size, and velocity dispersion are estimated for the dark matter halo associated with each tentative group; then galaxies are then re-assigned using halo information; the steps described above are reiterated until the group memberships converge. The final halo masses are obtained using the halo mass function given by \cite{Warren2006halomass}. Satellite galaxies located within $R_{180}$ of a halo and have stellar masses between [$10^{10}M_\odot$, $10^{11.5}M_\odot$] are used. Under these restrictions, our final sample includes 457 galaxies from a total of 115 halos (halo mass range [$10^{11.6}M_\odot$, $10^{14.7}M_\odot$]) among which, there are 54 star-forming galaxies.  

The following analysis will include several properties of galaxies and the derivation of our data is summarized below. Stellar masses, specific star formation rates (sSFR) and gas-phase metallicities are extracted from the MPA-JHU (Max-Planck Institute for Astrophysics - John Hopkins University) value-added galaxy catalogue of SDSS DR7 \citep{Kauffmann2003stellarmass}. Gas-phase metallicities can only be calculated for star-forming galaxies due to the emission limit of the OH line, so we only use star-forming galaxies to study gas-phase metallicity in the following analysis.  As for stellar metallicity, we use the catalogue based on SDSS DR2 \citep{Gallazzi2005metallicities}. The cross-matching of the radio data and SDSS DR2 excludes data from 4 pointings due to the incompleteness of SDSS DR2 compared to SDSS DR7. However, this exclusion has little effect on our final results. Under the same restrictions described above, the final sample for SDSS DR2 counterparts has 352 galaxies from 94 halos. We only use this sample for stellar metallicity analysis.


\section{Simulation data}
\label{sec:simd}

In this section, we outline the two semi-analytical galaxy formation models and two hydro-simulations used in this work. Table~\ref{Tab:1} lists the treatments of environmental effects, specifically ram-pressure stripping and tidal stripping in each semi-analytic model. 


\begin{table*}
    \footnotesize 
    \caption{ Table of ram-pressure stripping and tidal stripping mechanisms treatments in different SAMs. }
    \begin{tabular}{ccclcc}
    \hline
    \multicolumn{6}{c}{\textbf{Ram-pressure stripping}}  \\ \hline
    & G-X20GRA  & \multicolumn{2}{c}{G-X20RPS}& L-H15  & L-A21\\ 
    \hline
    Gas type & Hot gas  & \multicolumn{2}{c}{Hot gas and cold gas}  & Hot gas  & Hot gas \\ 
    \hline
    Limit    & satellite galaxies & \multicolumn{2}{c}{\begin{tabular}[c]{@{}c@{}}cold gas stripped only when \\ its radius is larger than hot gas;\\ satellite galaxies\end{tabular}} & \begin{tabular}[c]{@{}c@{}}Satellite galaxies within 1$R_{200}$ ;\\ halo mass $M_{200}>10^{14}M_\odot$;\end{tabular}  & All satellite and central galaxies \\ 
    \hline
    \multicolumn{6}{c}{\textbf{Tidal stripping}}\\ \hline
    Gas type & Hot gas  & \multicolumn{2}{c}{Hot gas}& Hot gas  & Hot gas   \\
    \hline
    Limit    & All FOF satellites     & \multicolumn{2}{c}{All FOF satellites}  & \begin{tabular}[c]{@{}c@{}}Satellites within 1$R_{200}$ of all halos\end{tabular} & \begin{tabular}[c]{@{}c@{}}All FOF satellites\end{tabular} \\ \hline
    \end{tabular}
    \label{Tab:1}
\end{table*}

\subsection{GAEA}

We select two versions of the model GAlaxy Evolution and Assembly (GAEA)  semi-analytic model 
from \citep[][hereafter X20]{xie2020influence}. X20 is developed on the basis of \citet{de2007hierarchical}, with a treatment for the non-instantaneous recycling of metals, energy and gas \citep{de2014elemental}, a modified stellar feedback method \citep{hirschmann2016galaxy}, an explicit modelling of H$_2$-based star formation and partitioning of HI and H$_2$ \citep[][hereafter X17]{xie2017h2}. X20 updated the treatment of environmental effects and hot gas and cold gas step-by-step and constructed three updated versions. Here we use two versions of the model called GRADHOT and RPSCOLD in X20. We refer to them in the following as G-X20GRA and G-X20RPS, respectively. 

The G-X20GRA model implements a treatment for tidal stripping and RPS to gradually remove hot gas from satellite galaxies, while the G-X20RPS model additionally includes RPS of cold gas. The tidal stripping of hot gas from satellite galaxies is assumed to be proportional to the stripping of dark matter from the parent dark matter substructures.
The RPS of hot gas and cold gas is computed by comparing the ram-pressure and gravity-binding pressure. Moreover, the G-X20RPS model assumes cold gas to be protected by a hot gas shell. Only cold gas outside the hot gas radius can be removed by RPS. X20 shows that both the G-X20GRA and G-X20RPS are in good agreement with quenched fractions measured for SDSS galaxies. 
The G-X20RPS can reproduce the gas scaling relations measured for central and satellite galaxies from xGASS \citep{Catinella2018xGASS} and xCOLDGASS \citep{Saintonge2017xCOLD}. GAEA is also able to reproduce the observed evolution of $z< 0.7$ stellar-phase metallicity and the evolution of gas-phase metallicity at $z< 3.5$ \citep{Fontanot2021evolution}, taking into account the uncertainties in observations.

Both models are based on the Millennium Simulation \citep[MS][]{springel2005simulations}, in which a WMAP1 cosmology is adopted ($\Omega_m=0.25$, $\Omega_\Lambda= 0.75$, $\Omega_b = 0.045$, $h = 0.73$, $\sigma_8 = 0.9$). The Millennium traces $2160^3$ particles from z = 127 to the present day. The box size is 685 Mpc with particle mass $1.2\times 10^9 $Mpc.

\subsection{L-GALAXIES}

We use two versions of the L-GALAXIES model, \citet[][hereafter L-H15]{henriques2015galaxy}  and \citet[][hereafter L-A21]{ayromlou2021galaxy} , based on the merger trees from the MS but re-scaled to the first-year Planck cosmology \citep{Planck16} with $\Omega_m=0.315$, $\Omega_\Lambda= 0.685$, $\Omega_b = 0.0487$, $h = 0.673$, $\sigma_8 = 0.829$. The re-scaled MS has a box size of $714$~Mpc and a particle mass of $1.43\times 10^9 M_{\odot}$.

L-H15 considers tidal stripping to gradually remove hot gas for all satellite galaxies within $R_{200}$. RPS of hot gas is also considered, but only for satellites in cluster halos $M_{200} > M^{14} M_{\odot}$.  
L-H15 shows good agreement with the observed red fractions up to $z\sim 3$, and the distribution of specific star formation rate at $z \sim 0$.

The work of L-A21 includes several improvements compared to L-H15, including updates described in \cite{henriques2020galaxies} (hereafter L-H20). L-H20 is updated for the treatment of gas cooling, the partition of HI and H$_2$, and H$_2$-based star formation in annuli \citep{fu2012}, as well as non-instantaneous chemical enrichment \citet{yates2013}. The re-calibrated L-H20 provides similar red fractions at $z=0$ and lower red fractions at higher redshifts compared to L-H15. 

Based on L-H20, L-A21 includes an updated approach to deal with RPS of hot gas in a more self-consistent manner \citep{ayromlou2019new}. 
The algorithm extracts a Local Background Environment (LBE) for each galaxy by analysing the dark matter particles. The ram pressure acting on each galaxy is measured from the of the LBE density and the relative velocity between the galaxy and its LBE.  
This version applies RPS to all galaxies, including central ones. The model quenched fractions match reasonably well the observations up to $z\sim 2$.

\subsection{EAGLE}
The EAGLE simulations \citep{schaye2015eagle, crain2015eagle} are a suite of hydro-dynamical simulations with cosmological parameters from the \cite{Planck14}. ($\Omega_m=0.307$, $\Omega_\Lambda= 0.693$, $\Omega_b = 0.04825$, $h = 0.6777$, $\sigma_8 = 0.8288$). This simulation is run with a modified version of the G\small{ADGET}-3 \normalsize TreeSPH code described by \citet{springel2005cosmological}. 
Dark matter halos are identified by the FOF algorithm and galaxies with the SUBFIND algorithm \citep{springel2001populating, dolag2009substructures}. In this work, we use the reference simulation with the largest volume named Ref-L100N1504. Its side length is 100cMpc.
The initial gas-particle mass is $1.81\times 10^6 M_\odot$ and the dark matter particle mass is $9.70\times 10^6 M_\odot$.

EAGLE reproduces the observed main sequence at $z =0.1$ \citep{schaye2015eagle}.  The quenched fraction of satellite galaxies in massive halos with $M_{h} < 10^{14} M_{\odot}$ is much higher than the quenched fractions measured for SDSS \citep{bahe2015}. \citet{Wang2018Dearth2}, however, found good agreement between EAGLE and SDSS for the quenched fractions against halo mass for central and satellite galaxies separately. The post-processed HI fractions of EAGLE galaxies in different environments are generally consistent with observations \citep{marasco2016environmental}. 
The scaling relations between stellar-phase/gas-phase metallicity and stellar mass of EAGLE agree well with observation for galaxies with $M_{\star} > 10^{10} M_{\odot}$. There are significant deviations for lower-mass galaxies.


\subsection{IllustrisTNG}

The IllustrisTNG project \citep[][hereafter TNG]{springel2018first, pillepich2018simulating, marinacci2018first, naiman2018first, nelson2018first} is a suite of cosmological hydrodynamical galaxy formation simulations run with the moving-mesh code  \texttt{AREPO} \citep{springel2010pur}. The initial cosmological parameters used in this simulation are Planck cosmological parameters ($\Omega_m=0.3089$, $\Omega_\Lambda= 0.6911$, $\Omega_b = 0.0486$, $h = 0.6774$, $\sigma_8 = 0.8159$). In this study, we use the publicly available data from TNG100, which consists of $1820^3$ dark matter particles and $1820^3$ baryonic particles in a 100cMpc box (in order to have a fair comparison with EAGLE, we don't use the 300cMpc box). The dark matter particle mass is $7.5\times 10^6 M_\odot$. The baryonic particle mass is $1.4\times 10^6 M_\odot$. 

The quenched fractions are in broad consistency with observational measurements \citep{Donnari2021quenched2}. The post-processed HI reasonably agrees with observational data from ALFALFA and xGASS \citep{Stevens2019}. 
TNG also reproduces observed gas-phase metallicity-stellar mass relations in the redshift range $0<z<2$ \citep{Torrey2019}.

\subsection{Galaxy Properties}

\begin{figure}
    \centering
    \includegraphics[width=0.47\textwidth]{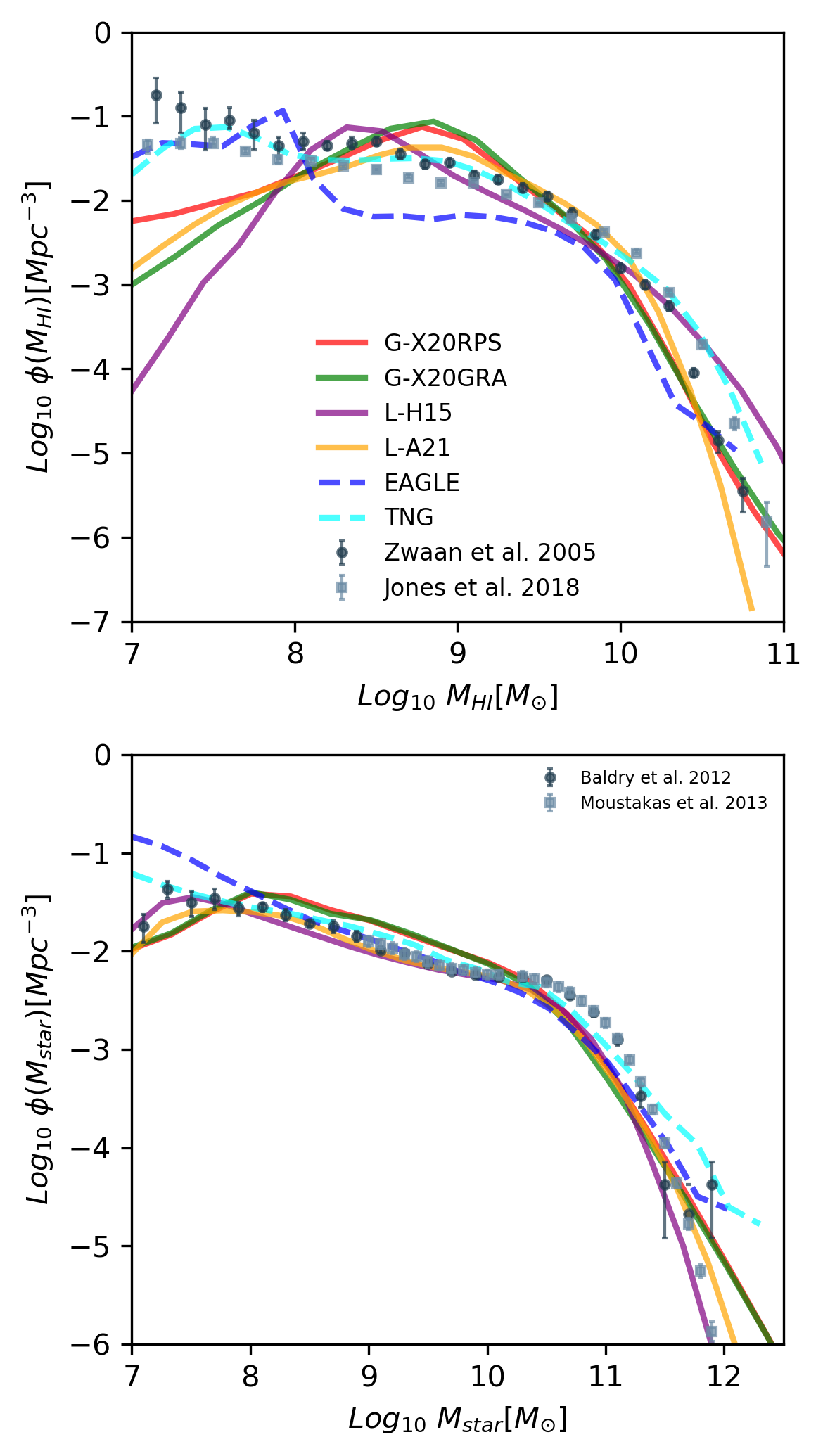}
    \caption{The top and bottom panels show the HI mass functions and the stellar mass functions at $z\sim 0$ for each model/simulation as labelled. 
    HI mass functions are compared to the observations by \citet{zwaan2005hipass} (circle) and \citet{Jones2018} (square). Stellar mass functions are compared to  \citet{baldry2012galaxy} (circle) and \citet{moustakas2013primus} (square). }
    \label{fig:MF}
\end{figure}

In this work, we investigate environmental effects on satellite galaxies at $z=0$. More specifically, we compare the HI mass, star formation rate (SFR), and metallicity of satellite galaxies as a function of the projected distance to their host halo centre with observation. 
For EAGLE, all properties are defined to be the sum of the properties of all star particles that belong to the corresponding subhalo and that are within a 3D aperture with radius 30 pkpc. 
For TNG, stellar mass, SFR and HI mass are calculated by total properties of all member particle/cells which are bound to this subhalo. Metallicities are calculated by all star or gas particles bound to the parent subhalo but restricted to gas cells within twice the stellar half mass radius. 

The stellar mass and SFR are direct outputs of the simulations, all assuming a \citet{Chabrier2003galactic} IMF. Galaxies may be characterized by extremely low SFR in hydro-dynamical simulations and SAMs, whose values are not well constrained in observational measurements. These measurements are often assigned a $\log$ sSFR around $-12$ \citep{Brinchmann2004physical} (sSFR is defined as SFR/$M_{\star}$). For consistency with this choice, we re-assign the sSFR of all model galaxies with $\log {\rm sSFR} < -12 yr^{-1}$ to random values in the range of $-12.5 < \log {\rm sSFR} < -11.5 yr^{-1}$. The HI masses and metallicities are computed differently for each model and simulation, as described below.

\begin{itemize}
    \item \textbf{HI mass}: G-X20GRA, G-X20RPS, and L-A21 implement an explicit treatment to partition cold gas into HI and H$_2$. 
    G-X20RPS and G-X20GRA use the empirical relation from \cite{blitz2006role} to account for the ratio between HI and H$_2$. In L-A21, the H$_2$ fraction is modelled using the fitting equations of \cite{mckee2009atomic} originated from \cite{krumholz2009atomic}. 
    EAGLE and TNG provide predictions for post-processed HI.
    In EAGLE, HI mass is calculated in each gas cell \citep{rahmati2013evolution} using the empirical observational relation \citep{blitz2006role}, in which the fraction of HI depends on the mid-plane pressure of this cell. For TNG, the detailed calculation of HI mass is given by \citet{diemer2018modeling} and \citet{diemer2019atomic}. They provide several methods of partitioning HI and H$_2$. Here we choose the model named GD14 \citep{gnedin2014line}. 
    L-H15 does not provide estimates for HI mass. We use a constant HI gas ratio $M_{HI}/M_{coldgas} = 54\%$ as in \cite{henriques2015galaxy}. Here we assume 75 per cent of cold gas is neutral hydrogen, and 70 per cent of neutral hydrogen is in the atomic phase.

    Figure~\ref{fig:MF} shows the HI and stellar mass functions of all hydro-dynamical simulations and SAMs at $z=0$. All hydro-dynamical simulations and SAMs match well the stellar mass function, with the exception of the massive end.
    Here, however, observational measurements are affected by large uncertainties due to small number statistics/volumes. 
    For the HI mass functions, TNG is in good agreement with observational measurements. SAMs coupled with MS can only resolve galaxies with $M_{\rm HI} > 10^9 M_{\odot}$ due to limited resolution. EAGLE predicts a lower HI mass function over the entire mass range, with a bump at around $M_{HI}\sim 10^8M_\odot$. According to \cite{crain2016eagle}, this corresponds to galaxies having no stellar particles, and is a result of numerical artificial effects.

    \item \textbf{Metallicity}: The stellar metallicity is presented as $Z_{\star}/Z_{\odot}$. For SAMs, $Z_{\star}$ is the total mass of all elements heavier than Helium divided by the total stellar mass.
    For EAGLE and TNG, the stellar metallicity is the total abundance of several elements. 
    The solar metallicity is $Z_\odot=0.02$. 
    
    Gas-phase metallicity is represented by oxygen abundance $12+$Log$_{10}$(O/H), where O/H is the number ratio of the two particles. G-X20RPS, G-X20GRA and L-A21 take into account the yields of individual elements, and output masses of oxygen and hydrogen directly. For L-H15,  we use the approximation in \citet{ma2016origin}, which assumes that oxygen abundance scales linearly with the gas mass metallicity (that is, $12+$Log$_{10}$(O/H)=Log$_{10}(Z_{gas}/Z_\odot)$+9. Here $Z_{gas}$ has the same definition as $Z_{\star}$ but in the gas phase. Both EAGLE and TNG provide separate Oxygen and Hydrogen mass fractions for the galaxies in the star-forming gas phase in the download catalogue.  
    
    In the observations, the gas-phase metallicity measurements are based on strong emission lines that are only visible for star-forming galaxies. In order to make a fair comparison, we also select star-forming galaxies with sSFR>$10^{-11} yr^{-1}$ in the analysis of gas-phase metallicity.

    \item \textbf{Projected distance}: For consistency, we measure projected distances between selected galaxies and the centre of each system in hydro-dynamical simulations and SAMs. 
    In order to account for projection effects, we set x, y, and z directions as line-of-sight in turns. The  projection depth is $10 Mpc/h$ 
    The impact of projection is discussed in  Appendix~\ref{sec:proj}.
    The projected distances are normalized by $R_{200}$,  the halo radius where the average enclosed density is 200 times the critical density of the universe. For the observational data, which use  $R_{180}$ as the normalization radius (defined as the halo radius with an average enclosed density 180 times the critical density), we convert it to $R_{200}$ using $R_{180}\approx 1.05 R_{200}$ \citep{Reiprich2013}.

\end{itemize}


\subsection{Galaxy sample selection}\label{sec:sc}

From each hydro-dynamical simulations and SAMs, we select halos with $10^{12.8}<M_{200}/M_\odot<10^{14.7}$. For each halo, we select all galaxies with a projected distance to the central galaxy less than 1.1 $R_{200}$. 
Following the observational sample, the selection criteria for galaxies from hydro-dynamical simulations and SAMs are $10^{10}<M_{star}/M_\odot <10^{11.5}$ at $z=0$.
Tab~\ref{Tab:2} lists sample sizes of all SAMs, hydro-dynamical simulations, and observations.

\begin{figure}
    \centering
    \includegraphics[width=\columnwidth]{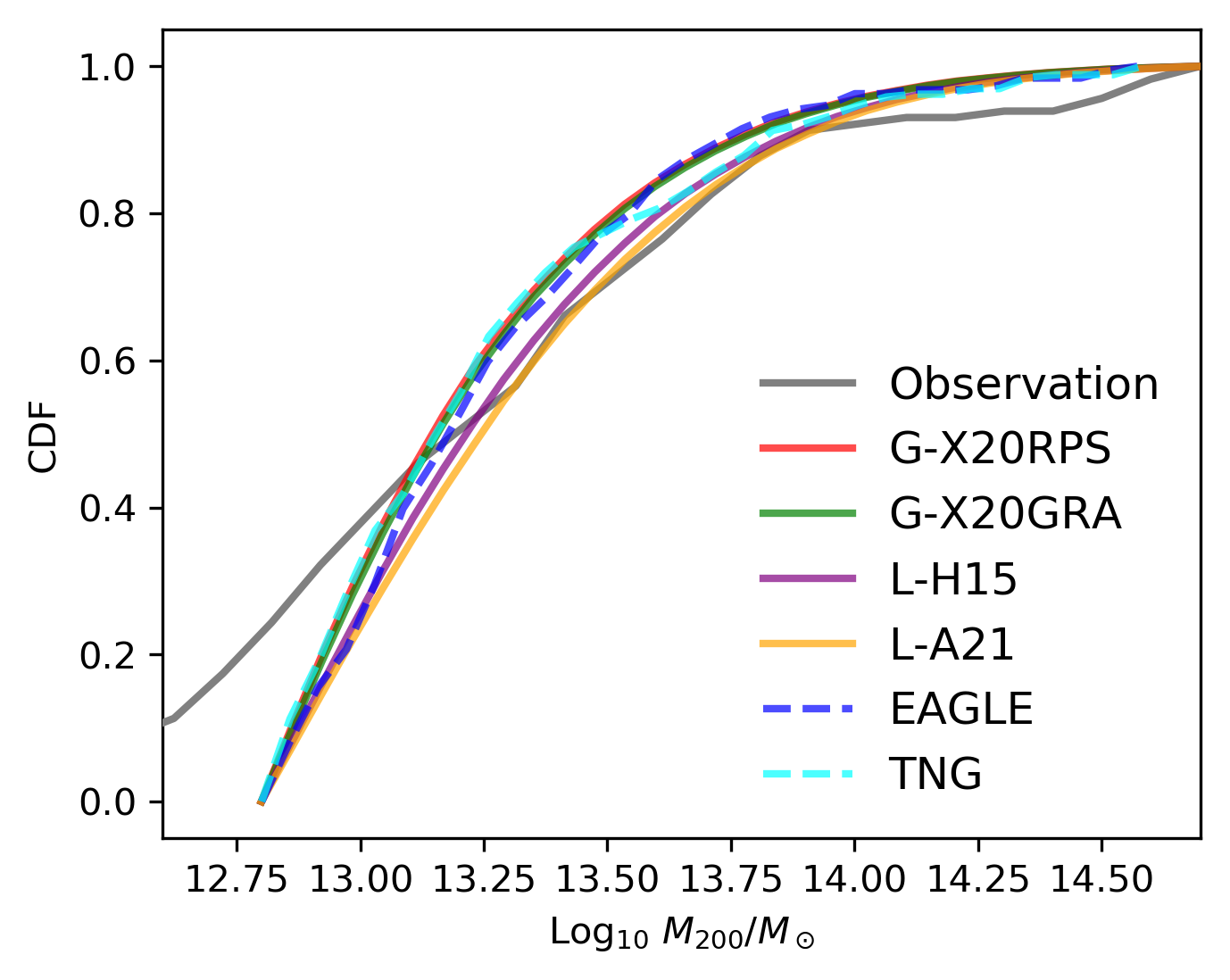}
    \caption{Cumulative distribution functions (CDFs) of host halo mass. The grey curve represents the observational sample, while the coloured curves represent hydro-dynamical simulations and SAMs. The y-axis value corresponds to the fraction of halos with mass smaller than $M_{200}$.
    }
    \label{fig:halodis}
\end{figure}

Figure~\ref{fig:halodis} illustrates the comparison between the halo mass distribution of the observational sample and the distribution obtained from hydro-dynamical simulations and SAMs. The observed galaxies have a minimum host halo mass of $10^{11.6}M_\odot$. However, in order to maintain consistency within the halo mass range  $M_{200} > 10^{13} M_{\odot}$, we have chosen a higher halo mass cut of $M_{200} > 10^{12.8} M_{\odot}$ for the model galaxies. The selected samples from hydro-dynamical simulations and SAMs display a slightly lower fraction of massive cluster halos than the observational sample. We will discuss the halo mass dependence in Sec~\ref{subsec:halomassdependence}.

\begin{table*}
    \scriptsize
    \fontsize{8}{8}\selectfont
    \centering
    \caption{Number of galaxies and halos in each selected samples. Rows marked with projected/3D are selected using projected/3D distances within 1.1 $R_{200}$. Galaxies are projected along the x,y, and z directions in turns. Therefore, the sample size is tripled when selected by projected distances.}
    \begin{tabular}{|c|c|c|c|c|c|c|c|c|c|c|}
    \hline
    & G-X20RPS & G-X20GRA & L-H15 & L-A21  & EAGLE & TNG & observation          \\ 
    \hline
    \begin{tabular}[c]{@{}c@{}}Galaxy number\end{tabular}            &1034955      & 965878     & 652352     & 505027       & 2557  & 4712   & 457 \\ \hline
    \begin{tabular}[c]{@{}c@{}}Star-forming \\ galaxy number\end{tabular}   &691788         & 735318             & 288059     & 189272       & 993  & 1530  & 54  \\
    \hline
    \begin{tabular}[c]{@{}c@{}}Star-forming \\ galaxy fraction\end{tabular}    &0.668    & 0.761  & 0.441      & 0.375          & 0.388  & 0.325    &         0.118           \\ 
    \hline
    Halo number  &199905   & 193632   & 188242     & 175973          & 565  & 792   & 115      \\ \hline
    \end{tabular}
    \label{Tab:2}
\end{table*}

\begin{figure*}
    \centering
    \includegraphics[width=1\textwidth]{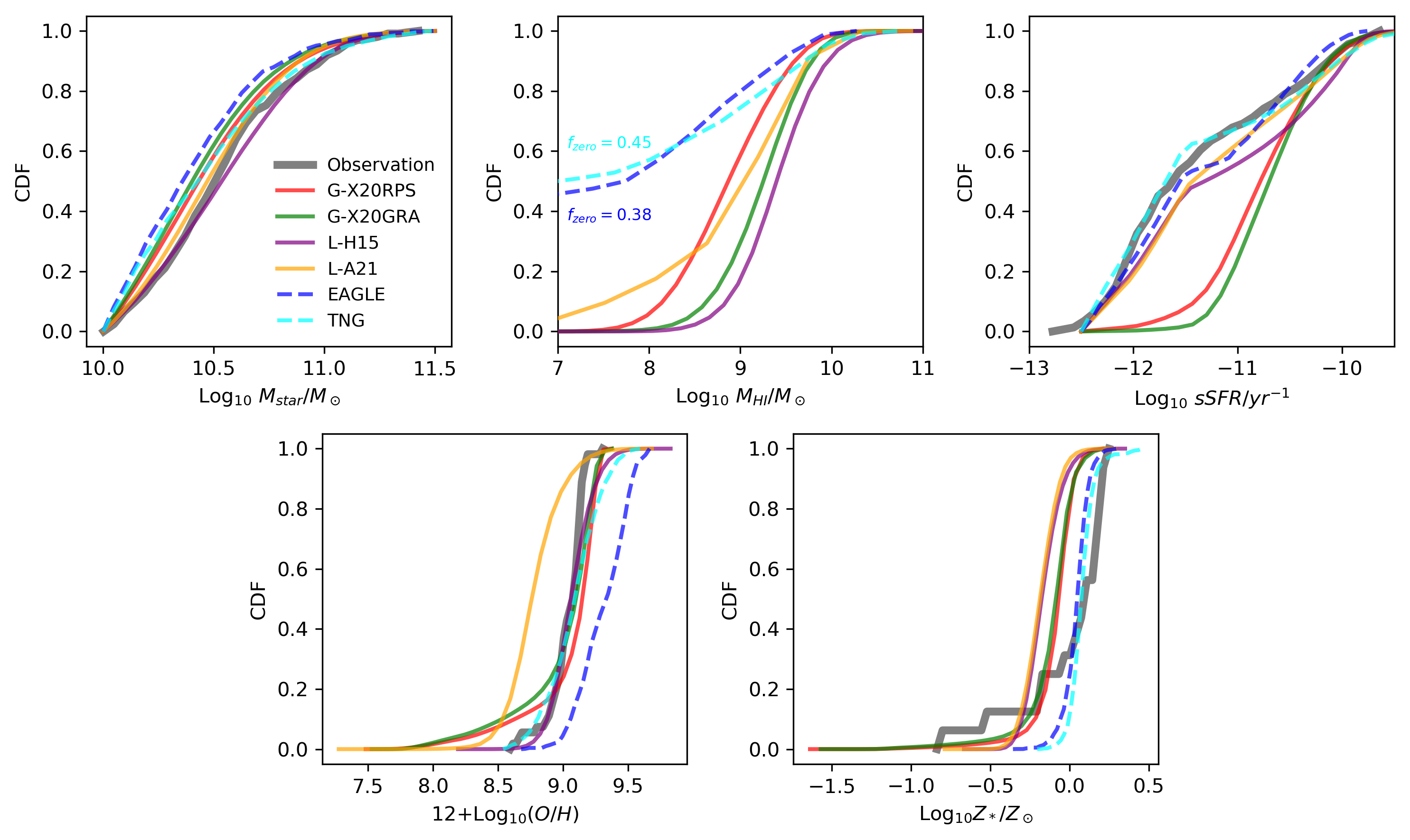}
    \caption{CDFs of stellar mass (upper left), HI mass (upper middle) and sSFR (upper right) as well as of gas-phase metallicity (lower left) and stellar metallicity (lower right). Note that the panel of HI mass does not include the observational sample as the spectral stacking technique used to estimate the observed HI mass does not give HI mass information for individual galaxies. 
    In the panel of HI mass distribution, the coloured numbers represent the fractions of galaxies with zero HI mass in the TNG and EAGLE, respectively. In the panel of sSFR, galaxies in SAMs/hydro-dynamical simulations with $ sSFR < 10^{-12} yr^{-1}$ are assigned random values between $[10^{-12.5},10^{-11.5}] yr^{-1}$.
    The fractions of galaxies with no star formation are: L-H15: 11.42\%; L-A21: 22.46\%; EAGLE: 44.94\%; TNG: 54.50\%.
    The two GAEA versions don't include any zero SFR galaxies. The lower left plot of gas metallicity only displays the distribution of star-forming galaxies (sSFR$>10^{-11}yr^{-1}$).}
    \label{fig:stdis}
\end{figure*}

Figure~\ref{fig:stdis} shows the cumulative distribution function (CDF) of galaxy properties for selected samples. 
The scatters between different hydro-dynamical simulations and SAMs are notable. The observational results fall within the scatter range. In general, galaxies selected from hydro-dynamical simulations and SAMs are less massive, more star-forming, and less enriched than those in the observational sample. 
It is worth noting that the observational sample we used is biased to lower SFR, compared to the SDSS. 
The two GAEA samples show higher sSFR distribution than other samples because the quenched cloud in GAEA has a peak sSFR located at $\sim 10^{-11} yr^{-1}$ \citep{xie2020influence}.  

The scatter between hydro-dynamical simulations and SAMs is even larger for the HI mass distribution. Generally, galaxies selected from semi-analytic models have more HI than galaxy samples from hydro-dynamical simulations. 
In EAGLE and TNG, there are $\sim 38$ per cent and $\sim 45$ per cent of galaxies contain no HI at all. 
In semi-analytic models, however, even quenched galaxies can have a HI mass that is larger than $\sim 10^8 M_{\odot}$ in L-H15 models and G-X20GRA models. Galaxies in the G-X20RPS and the L-A21 models have HI masses down to $\sim 10^7 M_{\odot}$. The HI mass distribution of the observational sample is not shown in this panel, as we use the stacked HI mass within a specific region, thus lack the HI mass for individual galaxies. We understand that the observed galaxies with extremely low HI mass are also included in the analysis. Therefore we also include simulated galaxies with no HI content.

\section{Results}\label{sec:re}

In this section, we compare the entire model selected samples with observation, then divide them by host halo mass into two groups to investigate the halo mass dependence. Additionally, we extend the previous observational survey to explore the metallicity trends. Given the large scatter across various samples, we focus on the trend as a function of projected distances, rather than on the normalization. 

\subsection{HI mass and star formation rate}

\begin{figure*}
    \centering
    \includegraphics[width=0.9\textwidth]{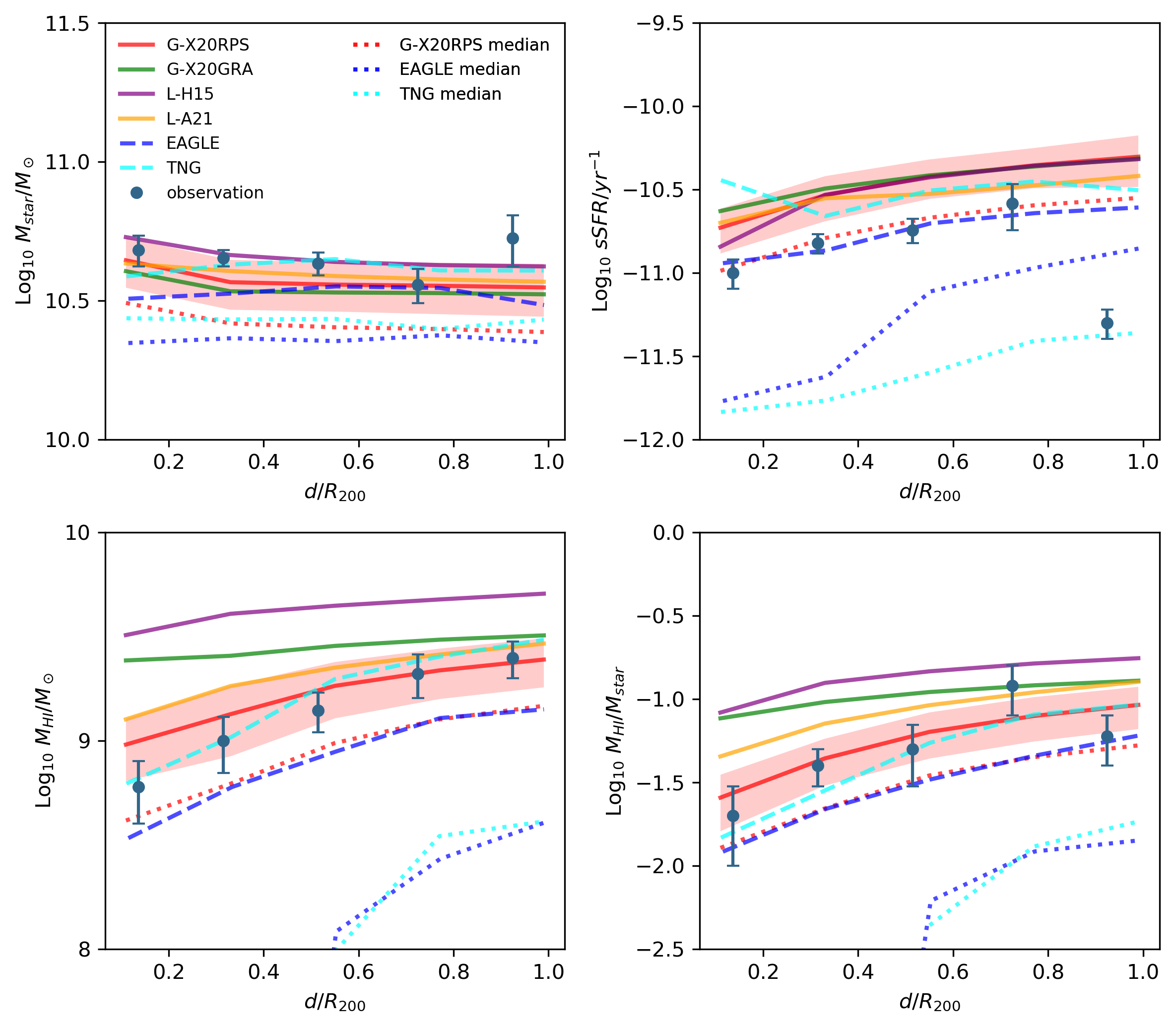}
    \caption{Radial distribution of averaged stellar mass (upper left), sSFR (upper right), HI mass (lower left) and HI mass function (lower right) and their comparison to the observation (symbols). Solid Lines and dashed lines represent the average values using the jackknife resampling method, the same as the observation. The shaded area denotes errors and displays only for model G-X20RPS for clarity. To avoid excessive lines, we opt to show only the median distribution as dotted lines for G-X20RPS, EAGLE, and TNG. Note that the bin at the largest distance of the observational sample should be interpreted with caution due to its small sample size \citep{hu2021atomic}. For example, the lowest value in the last bin of sSFR is caused by galaxies with highest averaged stellar mass.}
    \label{fig:stra}
\end{figure*}

The observational work of \cite{hu2021atomic} has confirmed a decreasing trend of HI mass for satellite galaxies from the outside to the halo centre. Here we discuss whether the trend exists in hydro-dynamical simulations and SAMs, and how this depends on the star formation activity of galaxies, as shown in Figure~\ref{fig:stra}.   
For consistency with observation, we plot the logarithmic mean value calculated by using jackknife resampling method for SAMs/hydro-dynamical simulations. Additionally, we also plot the median values, which are less biased and better reflect the intrinsic trends than the logarithmic mean values. We plot the median properties for only three SAMs, G-X20RPS, EAGLE, and TNG, to avoid overcrowding the figure with too many lines.

The HI mass and star formation rate correlate primarily to stellar mass. Therefore it is important to check if there's any systematic bias between the stellar mass of observations and simulations at given projected halo-centric distances. 
The upper left panel of Figure~\ref{fig:stra} displays the mean stellar masses of galaxies at fixed radial distances. The differences between the observation and simulations are small. 
Also, the mean stellar mass of the observational sample does not depend on the radial distance of galaxies significantly. The same trend is found for all hydro-dynamical simulations and SAMs.

\begin{figure*}
    \centering
    \includegraphics[width=\textwidth]{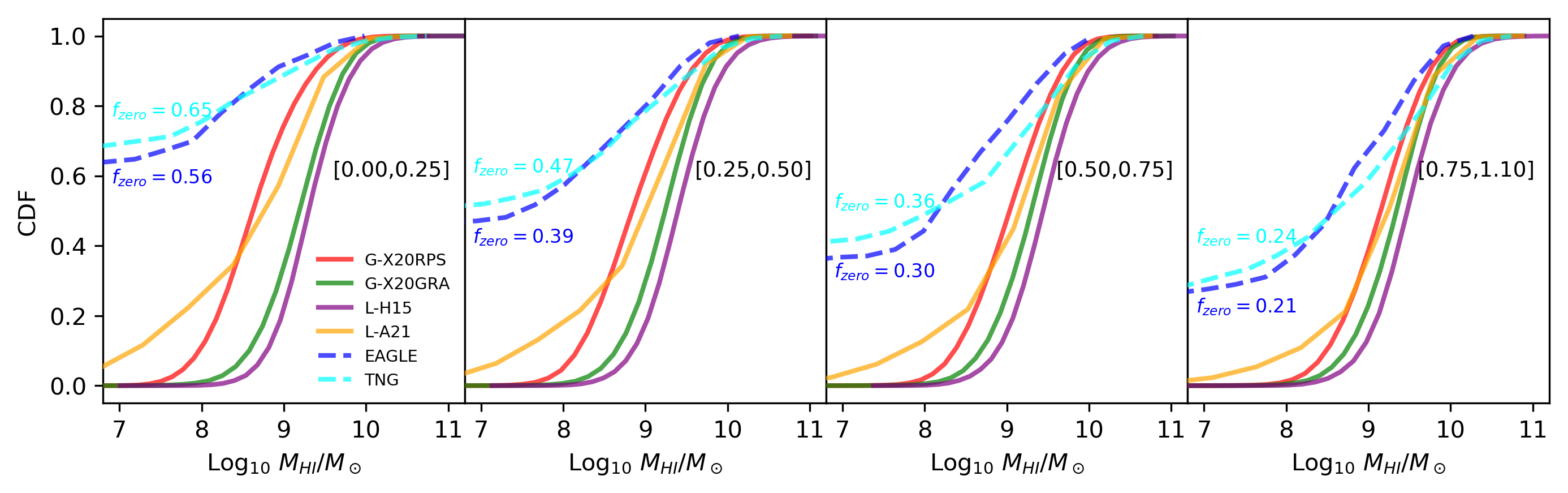}
    \caption{CDFs of HI mass in different normalized radial projected distance bins. Four panels represent ranges [0.00,0.25], [0.25, 0.50], [0.50, 0.75], [0.75, 1.00], respectively. As in Figure~\ref{fig:stdis}, the coloured percentage represents the fraction of galaxies with zero HI mass of the hydro-dynamical simulations TNG and EAGLE. }
    \label{fig:HIdira}
\end{figure*}

The bottom two panels of Figure~\ref{fig:stra} show the HI masses and HI mass fractions as a function of their distance to the halo centre. 
All hydro-dynamical simulations and SAMs predict decreasing HI mass and HI mass fraction with decreasing distances, which is consistent with observations. The decline of HI mass and fractions becomes steeper in the inner regions, indicating a stronger impact of the environment in the halo centre. 

Comparing the G-X20RPS and G-X20GRA, the former sample shows a steeper decline of HI than the latter. The RPS of cold gas is essential to reproduce the evolution trend observed in data in the GAEA model.  Figure~\ref{fig:HIdira} shows CDFs of HI mass in narrow bins of projected distances.
With the additional implementation of RPS on cold gas, the G-X20RPS sample includes a notably larger fraction of low HI mass galaxies than the G-X20GRA. The difference becomes more obvious at smaller distances.

In the case of L-Galaxies, the L-A21 sample is in better agreement with the observation than L-H15 in Figure~\ref{fig:stra}.
Both models do not consider the RPS of cold gas. The decrease in HI is a reflection of the decrease of hot gas and hence of the suppression of gas cooling. 
\citet{ayromlou2021galaxy} found that the hot gas to stellar mass ratio is lower in the L-A21 for $\sim 10^{10.5} M_{\odot}$ galaxies in group halos ($\sim 10^{13.5} M_{\odot}$), with respect to L-H15. As a result, galaxies from L-A21 retain less cold gas than those from L-H15.  
As shown in Figure~\ref{fig:HIdira}, the L-A21 sample includes a non-negligible fraction of HI-poor galaxies at $M_{HI} < 10^8 M_{\odot}$. This fraction increases towards the halo centre, where such HI-poor galaxies are not found in the L-H15. 

Both the RPS on cold gas in G-X20RPS and the updated treatment of RPS stripping on hot gas in L-A21 help to improve the model predictions with respect to previous renditions of the same physical models.

As for hydro-dynamical simulations, both EAGLE and TNG predict a declining trend for HI mass and fractions that is similar to the observations (see Figure~\ref{fig:stra}). 
In Figure~\ref{fig:HIdira}, the cumulative HI mass distributions of galaxies in TNG and EAGLE are biased towards low HI masses, compared to the semi-analytic models. 
Even at the virial radius, more than 10 per cent of galaxies in hydro-simulations have no HI. Moving closer to the halo centre, more galaxies in hydro-simulations have zero HI mass. This indicates a strong suppression of HI in EAGLE and TNG due to environmental effects. 

The logarithmic mean values in Figure~\ref{fig:stra} are dominated by the HI-rich galaxies, because of the non-normal distribution of HI mass. We also include the median values to provide a less biased trend. Interestingly, while the average HI masses of TNG and EAGLE don't significantly differ from the semi-analytic model G-X20RPS, their median values are offset significantly below the average trends. In the EAGLE and TNG simulations, the median HI fractions remain below $10^{-1.5}$ even at the virial radius. The fact that the mean and median are so different is a consequence of having a lot of galaxies with zero HI mass, as noticed in Figure~\ref{fig:HIdira}.
The suppression of HI for galaxies in dense environments is much stronger in hydro-simulations than in semi-analytic models. Surveys for HI-poor galaxies around cluster halos may provide more constraints on the environmental quenching physics.

The sSFR as a function of the projected radius is shown in the upper-right panel of Figure~\ref{fig:stra}. All hydro-dynamical simulations and SAMs except EAGLE samples have higher or comparable average sSFRs than observational samples at all distances. As mentioned before, the observational sample is biased to lower sSFR galaxies.   
For the observational sample, the average sSFR decreases from $10^{-10.6} yr^{-1}$ at $0.7R_{200}$ to $10^{-11} yr^{-1}$ at $0.2 R_{200}$. 
All hydro-dynamical simulations and SAMs show a similar decreasing trend with the average sSFR declining $\sim 0.4$ dex from $1 R_{200}$ to $0.25 R_{200}$. 

By comparing the G-X20RPS sample with the G-X20GRA sample, we find that the inclusion of RPS on cold gas slightly amplifies quenching effects within the $0.5 R_{200}$. 
It can also be noticed that the updated RPS approach implemented on L-A21, which considers the effect of ram-pressure stripping on central galaxies, intrigues a suppression of star formation before galaxies fall into the host halos compared to L-H15. The difference between these two models can be witnessed at $ R_{200}$.

The radial distribution of average sSFR for the EAGLE sample matches very well the observations. In addition, its median sSFR drops significantly at around $0.5 R_{200}$, as a consequence of the rapid transition of EAGLE galaxies into a quenched phase at this radius.
In the case of the TNG sample, the average sSFR of galaxies decreases from the $1 R_{200}$ to $0.3 R_{200}$, then increases in the innermost bin. The median sSFR exhibits a continuous decrease from $R_{200}$ to the innermost bin. We have checked that the innermost average sSFR is biased by a few galaxies with very high star-formation rates in the halo centre. 
Interestingly, the majority of galaxies in the TNG sample at the virial radius are quenched, when half of the galaxies still have a large fraction of HI $M_{HI}>10^{9} M_{\odot}$ (Figure~\ref{fig:HIdira}).

\subsection{Halo mass dependence}
\label{subsec:halomassdependence}

\begin{figure*}
    \centering
    \includegraphics[width=1\textwidth]{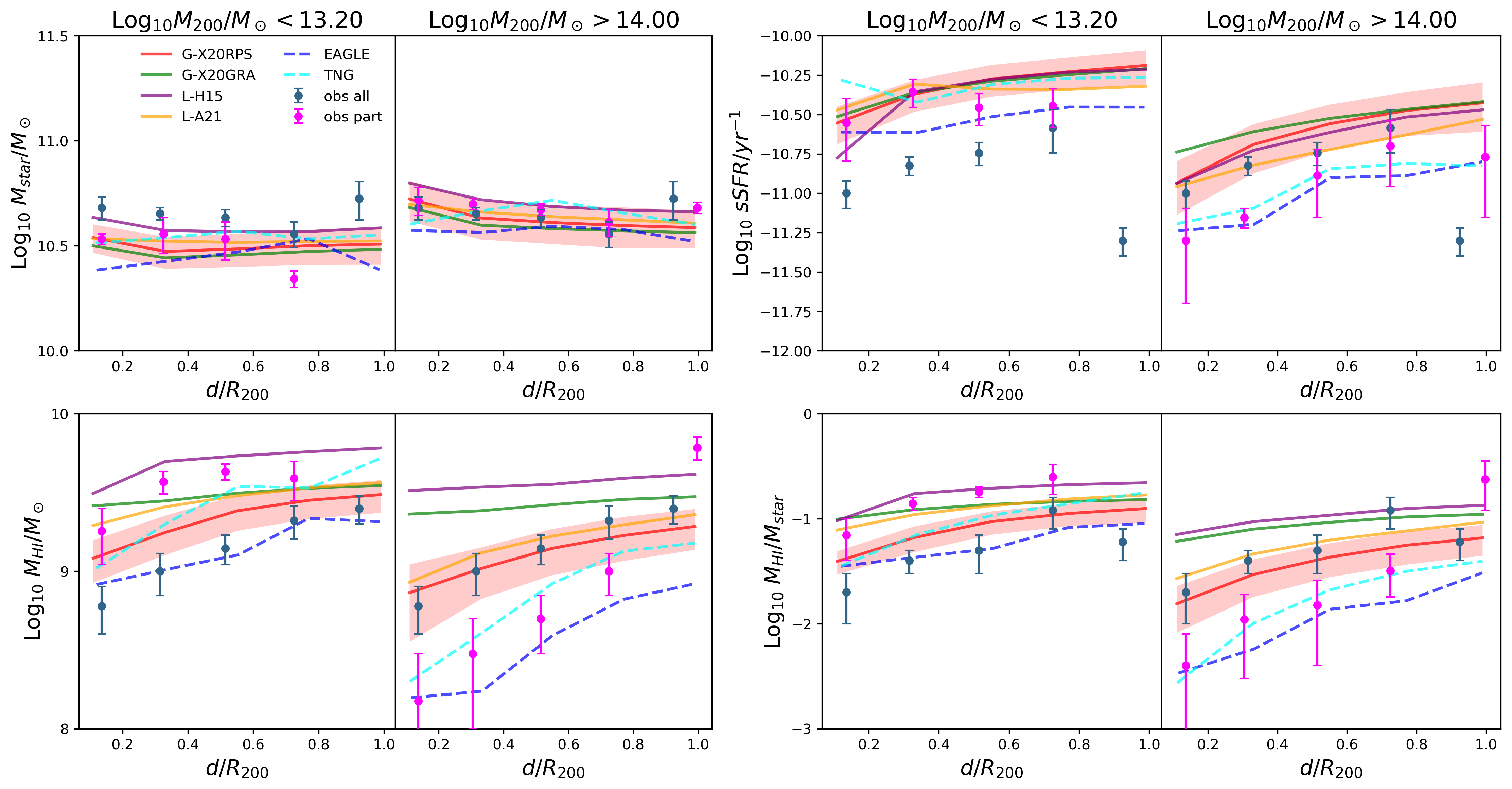}
    \caption{Same as Figure~\ref{fig:stra} but in different host halo mass range. From left to right, top to bottom, the paired panels show the radial projected distance distribution of stellar mass, sSFR, HI mass, and HI mass fraction. For each paired panel, the left one shows the results of lower-mass halos with $M_{200}/M_\odot<10^{13.2}$. The right shows results of more massive halos with $M_{200}/M_\odot>10^{14.0}$ for simulations as well as the observation.
    The magenta circles represent the result for observed galaxies in the given halo mass bin. We also include the results for all observed galaxies as blue circles for reference.  Note that the fifth bin (if plotted) of the observation is unreliable due to its small sample size. 
    }
    \label{fig:hmassbinra13_14}
\end{figure*}

It is widely accepted that environmental effects depend on halo mass. In this section, we split galaxies into two groups based on their host halo mass. Specifically, we select galaxies in low mass halos with $M_{200}/M_\odot<10^{13.2}$ and cluster halos with $M_{200}/M_\odot>10^{14.0}$. The results are shown in Figure~\ref{fig:hmassbinra13_14}. Note that hydro-simulations, especially EAGLE, include fewer massive cluster halos than SAMs and observational data, making their predictions more affected by Poisson noise.

The average HI mass and fractions present a clear dependence on the host halo mass across all the samples. Generally, galaxies residing in massive halos tend to have lower HI mass and fractions compared to those in low-mass halos. These differences already exist in the outermost bin.
At $\sim 0.7 R_{200}$ in the observation, the HI fraction in massive halos is $\sim 0.8$ dex lower than in low mass halos. This decrease is $\sim 0.2$ dex for all SAMs and $\sim 0.4$ dex for the two hydro-simulations. Therefore, EAGLE and TNG can better reproduce this decline.

The declining rate of HI also varies depending on the host halo mass. Observed galaxies in the innermost bin have $\sim 0.6$ dex lower HI mass than galaxies at distances $0.6 < d/ R_{200} < 0.8$ in low-mass halos. This difference becomes $\sim 1$ dex in massive halos. 
This halo mass dependence pattern is only evident in G-X20RPS, L-A21, EAGLE, and TNG. Hydro-simulations better capture the influence of halo mass on HI content. TNG yields the closest agreement with observational results in both low-mass and high-mass halos. EAGLE reproduces well the data for the most massive haloes, but the predicted HI mass decline towards the centre is milder and its overall HI content is considerably lower than the observations in low-mass halos.
In the framework of G-X20RPS and L-A21, satellite galaxies in massive halos show higher HI fractions and a milder decrease from the halo edge to the halo centre compared to the observational sample, indicating that the depletion of HI in SAMs are not as efficient as observations suggest. 
For satellites in low-mass halos, G-X20RPS predicts a lower HI fraction but a similar observed decreasing pattern from the halo outskirts to the centre, while L-A21 is closer to the observation with a milder decrease.
In both cases, L-A21 is $\sim 0.1$ dex higher than G-X20RPS.

The average sSFR shows similar halo mass dependences. The overall observed sSFR is higher with a milder declining rate in low-mass halos than in more massive halos.  
The same tendency also holds true for most hydro-dynamical simulations and SAMs. This means that satellite galaxies located within cluster halos tend to exhibit reduced star-formation activity and begin quenching at greater radii compared to their counterparts in lower-mass halos. The model L-A21 provides the closest agreement with the observational results for low-mass halos, while hydro-simulations show better consistency with the observation in high-mass halos.

It is well known that the environmental effects would deplete the diffuse HI earlier than the more dense star-forming gas. The quenching of satellite galaxies occurs later than the depletion of their HI gas. This `delayed quenching' is found for observed samples in both low-mass and high-mass halos. 

Focusing on G-X20RPS and G-X20GRA, the former model predicts lower HI fractions and sSFR along with steeper decreasing trends towards the halo centre in both low- and high-mass halos. However, in high-mass halos, the offset between these two models becomes larger, particularly towards the halo centre, indicating a stronger halo mass dependence in G-X20RPS. This means that RPS of cold gas in the GAEA model is more effective in depleting HI and quenching galaxies in high-mass halos, which is consistent with the observations.

L-A21 also shows lower HI fractions in low- and high-mass halos  than L-H15. The discrepancy between these two models also widens in higher-mass halos, suggesting that the updated stripping algorithm implemented in L-A21 more effectively depletes HI in high-mass halos. However, their sSFR comparison reveals a different behaviour. Though L-A21 generally predicts lower sSFR and is more consistent with the observation, L-H15 predicts a steeper declining trend in both low- and high-mass halos, providing a more accurate treatment of quenching rates in satellite galaxies.

Though SAMs do show some degree of halo mass dependence, this dependence is not as dramatic as in hydro-simulations. This might be because hydro-simulations have a better treatment of the spatial distribution of the gas components, resulting in stronger HI depletion in high-mass halo mass and stronger quenching mechanisms.

\subsection{Environmental effect on metallicity}

\begin{figure*}
    \centering
    \includegraphics[width=0.9\textwidth]{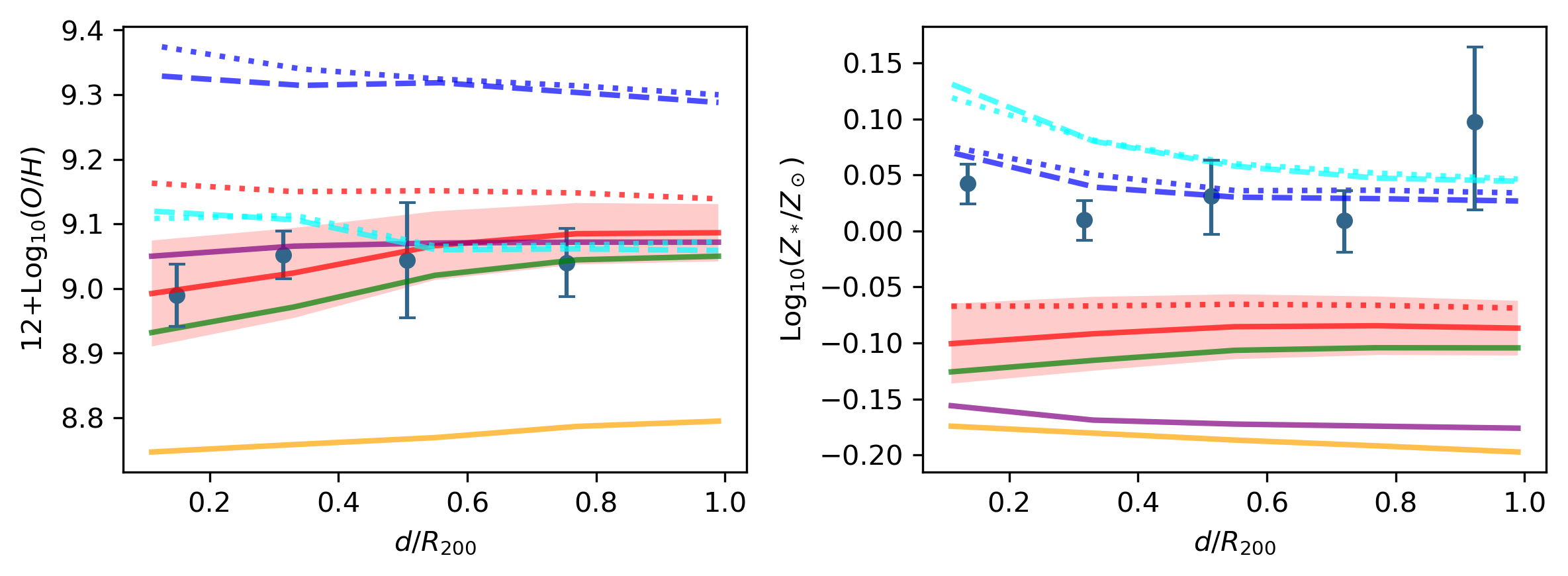}
    \caption{Radial projected distribution of oxygen abundance (left)  of star-forming galaxies and stellar metallicity (right) of all galaxies, compared to observations (symbols). The projected distance is normalized by $R_{200}$. As in Figure~\ref{fig:stra}, solid curves show mean results, and dotted lines show the median values. Here we use $Z_\odot=0.02$. Note that the fifth bin (if shown) of the observation is unreliable due to its small sample size.}
    \label{fig:metalra}
\end{figure*}

The gas-phase and stellar-phase metallicity depend on the local density surrounding the galaxy\citep{peng2015}. Except for the impact of assembly bias, environmental effects on galaxies are also reflected in metallicities. Star formation, gas accretion, and gas stripping could affect the gas-phase metallicity by reducing/increasing the gas fraction or injecting metals into the interstellar medium. The stellar metallicity reflects the evolution history of a galaxy on a longer time scale.  In this section, we compare the radial distribution of gas-phase and stellar-phase metallicities between the observed sample and simulations, as shown in Figure~\ref{fig:metalra}. Since in the MPA-JHU value-added galaxy catalogue, gas-phase metallicity is typically estimated using strong emission lines. For consistency with observations, in this section we consider only star-forming galaxies with $sSFR>10^{-11}yr^{-1}$ from hydro-dynamical simulations and SAMs.

We noticed that the gas-phase metallicity of L-A21 is $\sim$ 0.2 dex lower than the observations. This deviation agrees with Figure 6 of \cite{yates2021galaxies}, where for stellar masses above $10^{10}M_{\odot}$, the gas-phase metallicity typically falls within the range of 8.6 to 8.8, $\sim$0.2 dex lower than the previous L-Galaxies model.  Conversely, EAGLE predicts metallicities $\sim$0.2 dex larger than the observations, see also \cite{bahe2017origin}. 

Regarding stellar metallicity, there is  an underestimation of 0.1 or 0.2 dex for all SAMs compared to the observational data, though they are still within the 1 sigma scatter of the observation \citep{hirschmann2016galaxy,yates2021galaxies}. This discrepancy also arises from our sample selection with $M_{start}>10^{10}M_{\odot}$. 
Therefore, in the following, we focus on the radial trends, and pay less attention to the absolute metallicity values.

The average metallicity of the observational sample does not depend on halo-centric radii at large distances, with a small decrease observed in the innermost radial bin. Interestingly, the stellar-phase metallicity increases towards the halo centre, whereas the gas-phase metallicity decreases. We interpret the trend as a result of multiple reasons. By selecting star-forming galaxies, we tend to select galaxies that either suffer weaker environmental effects and remain star-forming or those outside of the host halos and are selected by projection. The average gas-phase metallicity could also be biased by a few low-metal star-forming galaxies. These influences are witnessed in the results of SAMs.

The evolution trend of the L-Galaxies series model are consistent with observation, despite the fact that galaxies in L-A21 are metal-poorer than the observational sample. Both L-H15 and L-A21 show decreasing gas-phase metallicity and increasing stellar metallicity towards halo centre. However, by tracing back the evolution of all satellite galaxies, we find increasing trend for both stellar and gas-phase metallicity. About half satellite galaxies in the halo centre, however, are excluded in the analysis of gas-phase metallicity because they are quenched. Therefore the average gas-phase metallicity decreases.

G-X20RPS and G-X20GRA show similar trends in the evolution of gas-phase metallicity, which is consistent with observational results. Their average stellar metallicity of the entire sample also decreases towards the halo centre. The decreasing trend is caused by satellite galaxies that have experienced recent gas cooling. A few satellite galaxies have a large fraction of hot gas at the time of infall which maintains gas cooling and star formation even when they are satellites. With the replenishment of metal-poor cooling gas, these satellite galaxies are relatively metal-poorer in both the gas-phase and stellar-phase than others with relatively lower star-formation rates. These gas-poor galaxies cause a slight downward in the mean metallicities. But the impact on median metallicities is negligible.
The median gas-phase metallicity and stellar-phase metallicity are flat or increasing towards the halo centre.
We notice that with the RPS of cold gas implemented, the average gas-phase metallicities of galaxies in G-X20RPS are higher than their analogues in G-X20GRA.

Both EAGLE and TNG predict increasing gas-phase metallicity and stellar-phase metallicity towards the halo centre. The median results are consistent with the average values. 
environmental dependencies of metallicity have been discussed in previous works. \citet{wang2023} found that satellite galaxies are more metal-enriched in more massive cluster halos, whereas central galaxies are more metal-poor in more massive halos. They explain the difference as a result of the influence of environmental effects on satellite galaxies, and AGN feedback on central galaxies. \citet{donnan2022} have discussed the dependence of gas-phase metallicities on environment for the TNG simulation. They interpret the dependences found as a combined result of halo assembly bias and enriched gas inflow in dense environments. We argue that environmental stripping could increase the metallicity more efficiently in hydro-dynamical simulations than in the semi-analytic models. As stripping occurs outside-in, it would  removes  metal-poor gas. In semi-analytic models, gas stripping does not affect its metallicity.

\section{Discussion and Conclusion}\label{sec:cc}

In this paper, we investigate the influence of environment by focusing on the cold gas content of satellite galaxies. In particular, we compare the observed radial distributions of HI mass/fraction, sSFR, gas-phase metallicity and stellar-phase metallicity with several theoretical models: including the hydro-dynamical simulations Illustris TNG and Eagle, and the semi-analytic models GAEA and L-Galaxies. Additionally, we investigate the impact of halo mass dependence on these properties. Our conclusions are summarized below.

\begin{itemize}
    \item All hydro-dynamical simulations and SAMs show a decreasing HI content from the halo outskirts to the halo centre. G-X20RPS in SAMs and two hydro-simulations agree better with the observations. The decreasing trends also depend on the host halo mass. Both EAGLE and TNG can reproduce the rapid decreasing of HI content in cluster halos, and the milder decreasing trends observed in lower mass halos. G-X20RPS and L-A21 from SAMs also present this different decreasing trend in high-mass and low-mass halos, but predict a milder decrease of HI fractions in cluster halos.

    \item  Halo mass dependence is also found in sSFR as a function of cluster-centric radii. For lower-mass halos, the average sSFR only decreases at the innermost bin. While in cluster halos, the decline starts since $R_{200}$. The same is found for all hydro-dynamical simulations and SAMs, among which, EAGLE and TNG show a better agreement with data.

     \item The declining trend of HI is stronger than the decline of sSFR and starts at larger radii, in both lower-mass and massive halos. The environmental effects have a more pronounced and earlier impact on the HI gas compared to their effect on star formation.

    \item We find decreasing gas-phase metallicity for star-forming galaxies and increasing stellar-phase metallicity from the halo outskirts to the halo centre. The satellite galaxies would have increasing metallicity when suffering environmental effects. However, the selection of star-forming galaxies excludes metal-rich quenched satellite galaxies in halo centre and leads to a decrease in the gas-phase metallicity in halo centre.

\end{itemize}

Our findings highlight the crucial role of cold gas stripping in shaping the evolutionary trajectory of satellite galaxies within a dense environment. In the framework of GAEA, RPS stands out as the key reason for reproducing the drop of HI masses after galaxies fall into massive halos. Also, the inclusion of sophisticated treatments of ram-pressure stripping for hot gas in L-A21 does improve the consistency with observational results with respect to L-H15. Both improvements are more obvious in high-mass halos. 

For EAGLE and TNG, the dramatic drop in the median HI mass, as well as the fact that half of satellite galaxies in halo central regions have no HI left, indicates much stronger suppression of the gaseous content for satellite galaxies in these hydro-dynamical simulations than that in semi-analytic models. Such a strong environmental effect might be, at least in part,  attributed to artificial limited resolution or strong internal feedback that prevents gas from cooling and enhances gas stripping in hydro-dynamical simulations. 
Our work demonstrates that increasing the observational sample of HI measurements in massive haloes can provide useful constraints on the strength of environmental processes and ram-pressure stripping in particular.

\section*{Acknowledgements}
This work is supported by the National Key R\&D Program of China ( 2022YFA1602901), the NSFC grant (Nos. 11988101, 11873051, 12125302, 11903043), CAS Project for Young Scientists in Basic Research Grant (No. YSBR-062), and the K.C. Wong Education Foundation. LZX acknowledges support from the National Natural Science Foundation of China (No. 12041302). MH acknowledges funding from the Swiss National Science Foundation (SNF) via a PRIMA Grant PR00P2 193577 “From cosmic dawn to high noon: the role of black holes for young galaxies”. We are grateful to the referee for their many helpful suggestions.

\section*{DATA AVAILABILITY}
An introduction to the GAEA code, as well as data file containing published model predictions, can be found at the GAEA webpage \url{https://sites.google.com/inaf.it/gaea/}.
The data underlying this article will be shared on reasonable request to the corresponding author.



\bibliographystyle{mnras}
\bibliography{mnras_template}



\appendix

\section{Projection Effect}\label{sec:proj}
\begin{figure*}
    \centering
    \includegraphics[width=1.0\textwidth]{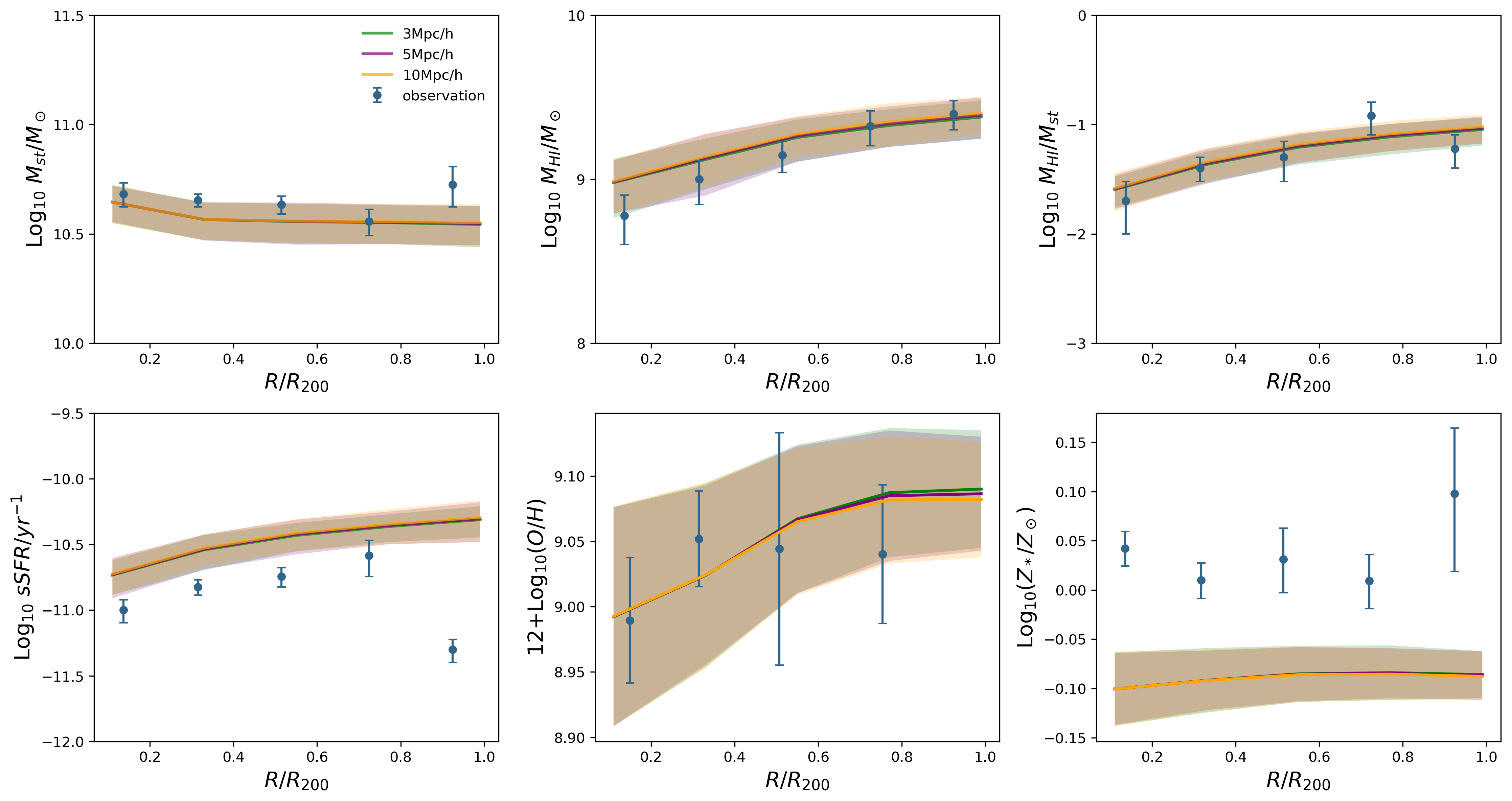}
    \caption{Radial properties of stellar mass, HI mass, HI mass fraction, sSFR,  the gas-phase metallicity of star-forming galaxies and stellar metallicity under different projection depth 3Mpc/h (green), 5Mpc/h (purple) and 10Mpc/h (orange). }
    \label{fig:proj}
\end{figure*}

Here we present the projection effect under different projection depths. We present its effect on the properties we discussed in this paper using G-X20RPS as an example, as shown in Figure~\ref{fig:proj}. The differences among project depths are negligible in all cases.


\label{lastpage}
\end{document}